\documentclass[aip, amsmath,amssymb, reprint]{revtex4-1}

\usepackage{graphicx}
\usepackage{dcolumn}
\usepackage{bm}
\usepackage[utf8]{inputenc}
\usepackage[T1]{fontenc}
\usepackage{mathptmx}
\usepackage{etoolbox}
\usepackage[dvipsnames]{xcolor}
\usepackage{url}

\makeatletter
\def\@email#1#2{%
 \endgroup
 \patchcmd{\titleblock@produce}
  {\frontmatter@RRAPformat}
  {\frontmatter@RRAPformat{\produce@RRAP{*#1\href{mailto:#2}{#2}}}\frontmatter@RRAPformat}
  {}{}
}%
\makeatother
\begin{document}

\preprint{AIP/123-QED}

% \title[Beyond memory-effect matrix-based imaging in scattering media by acousto-optic gating]{Beyond memory-effect matrix-based imaging in scattering media by acousto-optic gating}

\title{Beyond memory-effect matrix-based imaging in scattering media by acousto-optic gating}

\author{Elad Sunray}

 \affiliation{Institute of Applied Physics, Hebrew University of Jerusalem}

\author{Gil Weinberg}
\affiliation{Institute of Applied Physics, Hebrew University of Jerusalem}

\author{Moriya Rosenfeld}
  \affiliation{Institute of Applied Physics, Hebrew University of Jerusalem}

\author{Ori Katz}%
\affiliation{Institute of Applied Physics, Hebrew University of Jerusalem}
\email{orik@mail.huji.ac.il}

\date{\today}
             
\maketitle

\begin{quotation}
Imaging inside scattering media at optical resolution is a longstanding challenge affecting multiple fields, from bio-medicine to astronomy. In recent years, several groundbreaking techniques for imaging inside scattering media, in particular scattering-matrix based approaches, have shown great promise, but usually suffer from a restricted field of view (FOV) due to their reliance on the optical 'memory-effect'.
Here, we demonstrate that by combining acousto-optic spatial-gating with state-of-the-art matrix-based imaging techniques, diffraction-limited imaging beyond the optical memory-effect can be achieved in a robust fashion. Specifically, we show that this can be achieved by computational processing of scattered light fields captured under scanned acousto-optic modulation. 
The approach can be directly utilized whenever the ultrasound focus size is of the order of the memory-effect range, independently of the scattering angle.
\end{quotation}

\section{Introduction}

Imaging through scattering media presents a significant challenge, as the random scattering of optical waves leads to a loss of image resolution and contrast. This challenge is particularly critical in deep tissue imaging, where the ability to resolve microscopic features more than a fraction of a millimeter deep inside turbid tissues is currently extremely limited.

Optical imaging through scattering layers has improved in recent years \cite{bertolotti2022imaging}, as several methods for optical-resolution imaging behind scattering media have been demonstrated. Examples of these methods include speckle-correlation approaches\cite{bert12,katz14, edrei2016optical}, and coherent transmission- or reflection-matrix techniques\cite{kang17,lee22,najar23,haim2023image,zhang2023deep}. These approaches have demonstrated a remarkable efficiency in correcting distortions and scatterings in diverse settings, including soft-tissues \cite{kang17}, mouse skulls \cite{kwon2023computational}, and multicore fibers \cite{choi2022flexible}, among others.
However, these approaches often rely on exploiting the optical 'memory-effect,' or 'isoplanatism', which assumes the presence of correlations among scattered waves that originate from different points within the target. This requirement inherently restricts the field of view (FoV) of the above techniques, in the case of large scattering angles, since the scattered light fields from different points are mixed at the detection plane.

A potential solution to address the limited FoV, which applies to moderately-scattering media that scatter light to limited angles, involves illuminating separate isoplanatic regions in separate measurements, where in each measurement the illuminated area is contained within the isoplanatic patch \cite{gardner2019ptychographic,li2019image,zhou2020retrieval}. 
Alternatively, computational post-processing can be employed to separate these regions in the detection process rather than the illumination process \cite{trussell1978sectioned,alterman2021imaging,lee22,balondrade2023multi,najar23}. In these approaches, the imaged target is computationally divided into segments that are smaller than the isoplanatic patch, individually correcting each segment and stitching/mosaicking the results together to form a large FoV image. 
It is important to acknowledge, however, that in highly scattering media, where the scattering illumination/detection point-spread function (PSF) is spread over angles larger than the isoplanatic (memory-effect) angle, the different isoplanatic patches are mixed and overlap in the illumination/detection plane, respectively, making such 'mosaicking' approaches non-applicable.

To tackle this challenge, several computational approaches have been recently proposed to expand the FoV beyond the limitations imposed by the optical memory-effect. One approach includes examining higher singular vectors within a modified reflection matrix \cite{badon2020distortion}. However, this technique is not universally applicable. Another avenue involves modeling the scattering media as multiple scattering layers and subsequently correcting them \cite{haim2023image,kang2023tracing}. Nevertheless, these techniques face challenges in scalability with an increasing number of layers, as an increasing number of parameters (scattering layers, spatial phases, and amplitudes) need to be retrieved from a fixed number of measurements. 

On the other hand, a solution for imaging deeper inside a tissue, without reliance on the memory-effect is acousto-optic imaging/tomography (AOI) \cite{wang2004ultrasound,elson11,resink2012state}. AOI combines the advantages of ultrasound acoustic waves, which scatter much less than optical waves in soft tissues, and optical-contrast imaging.
In conventional AOI, a focused ultrasonic spot modulates the optical field at the target plane. The light passing through the acoustic focus experiences a frequency shift (commonly referred to as ultrasonic 'tagging'), allowing to separate it in detection from the rest of the light field. %We can obtain signals from the ultrasound focus by selectively filtering the modulated field in the detector plane and summing the signal intensity from this specific known area. 
By scanning the ultrasound focus over the target plane, and summing the acousto-optically modulated and detected signal at each ultrasound focus position, an optical-contrast image of targets embedded deep inside scattering samples is retrieved, albeit with a resolution dictated by the ultrasound focus dimensions, orders of magnitude larger than the optical resolution \cite{elson11}.

In recent years, significant efforts have been directed towards improving the resolution of acousto-optic tomography and overcoming the acoustic-diffraction limit in both imaging and focusing. In the case where a measurable amount of unscattered ballistic components exists in the light passing through the sample, images can be directly obtained from spatially-gated acousto-optic measurements \cite{jang2020deep,ko2023acousto}. 
In the diffusive regime, approaches based on super-resolution optical fluctuation imaging (SOFI) \cite{dertinger2009fast} construct acousto-optic images with improved resolution from the fluctuations of conventional acousto-optic measurements acquired under random unknown illuminations \cite{doktofsky20}. However, in practice, the large number of required measurements limit the resolution improvement to a factor of $\lessapprox 3$, far from the orders of magnitude improvement required to reach an optical diffraction-limited resolution.

Advanced acousto-optic techniques that exploit wavefront-shaping for coherently controlling the illumination have been shown to allow sub acoustic-diffraction focusing and imaging: techniques such as iterative phase-conjugation methods (I-TRUE) \cite{si2012breaking,ruan2014iterative}, time reversal of variance-encoded fields (TROVE) \cite{judkewitz2013speckle}, and decomposition of the acousto-optic transmission matrix (AOTM) \cite{katz2019controlling}, have all been shown to improve focusing resolution beyond the acoustic wavelength by effectively utilizing the first singular vector of the AOTM.
 
However, these advanced techniques have not been able to achieve optical-wavelength scale focusing, since the use of only the first singular-vector requires the number of optical modes contained within the acoustic focus (optical speckles) to be considerably smaller than that encountered in practice \cite{katz2019controlling}. 

Here, we demonstrate that the optical memory-effect can be leveraged to yield optical diffraction-limited imaging using the same datasets acquired in TROVE or AOTM experiments, by processing existing datasets by the state-of-the-art computational scattering compensation matrix-based techniques such as CLASS \cite{kang17,lee22} and distortion-matrix \cite{badon2020distortion}. This combination allows not only to provide optical diffraction-limited images, but also to reconstruct the scattering medium phase distortion. Most importantly, unlike TROVE or AOTM, the reconstruction convergence is independent on the number of optical speckle grains contained within the acoustic focus, and is possible whenever the acoustic focus size is of the order of the memory-effect or smaller.
By scanning the acoustic focus across the target plane, a large FoV reconstruction, well beyond the memory-effect, can be obtained by simply mosaicking the independent reconstructions, not requiring any assumptions on the scattering PSF.

Recently, a Ptychography-based speckle-correlation approach for diffraction-limited imaging beyond the memory-effect using acousto-optic tagging was presented \cite{rosenfeld21}. However, this approach suffers from several inherent limitations compared to our approach. These limitations originate from estimating both the autocorrelation from the measured speckle patterns and the ptychographic reconstruction algorithm requirements. Specifically, it strictly requires numerous jointly and highly overlapping scans of the target by the ultrasound focus, and it is sensitive to several heuristically-determined processing parameters \cite{rosenfeld21}. In addition, the correlography-based approach averages over the medium scattering and does not allow direct reconstruction of the scattering medium distortion, which is required for, e.g., optical focusing through the medium.

Below, we numerically and experimentally demonstrate that our proposed acousto-optic matrix-based approach can robustly reconstruct a wide FoV beyond the memory-effect, without requiring large overlaps between the different acoustic foci positions, exploiting the entire phase and amplitude information embedded within the acousto-optic holographic measurements. Moreover, we demonstrate that the matricial information can be acquired in a compressive manner by exploiting unknown random illuminations via the compressed time-reversal (CTR) matrix acquisition scheme known as CTR-CLASS \cite{lee22}.

%%%%%%%%%%%%%%% Figure 1 - Steup & Numerical %%%%%%%%%%%%%%%%%%%%%%
\begin{figure}[ht!]
\includegraphics{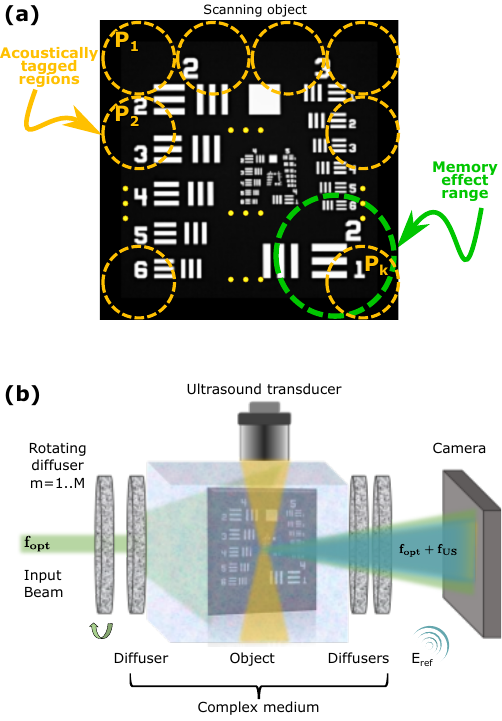}
\caption{\textbf{Imaging through scattering beyond the optical memory-effect by acoustic-optic tagging} (a) Our proposed method involves using an ultrasound probe to scan the object acoustically. The key aspect of this technique is using a probe with a focus size (depicted in yellow dashed circles) that is smaller than the isoplanatic patch size (illustrated in green dashed circle). This ensures that, after the necessary filtering, we obtain a field that has traversed exclusively through the area affected by acoustic modulation. By isolating this sonically modulated field, we can effectively apply algorithms that rely on the 'optical memory-effect'. (b) Our method utilizes a conventional holographic 'acousto-optic imaging' setup with a rotating diffuser, which generates uncorrelated speckle patterns on the object by a quasi-monochromatic laser beam (depicted in green) with a frequency centered around $f_{opt}$. While the speckles illuminate a large area, much larger than the optical memory-effect, by tagging a confined area on a target object using a focused ultrasound beam (shown in yellow) with a frequency of approximately $f_{US}$, we can extract the information from that confined area. This is repeated $M$ times, where each illumination occurs with the diffuser set to a different rotation angle. We extract the modulated field using holographic measurements by adding a reference beam with a frequency of $f_{opt} + f_{US}$.}
	\label{fig1}
\end{figure}

%%%%%%%%%%%%%%% Figure 2 - Numerical results %%%%%%%%%%%%%%%%%%%%%%
\begin{figure*}[t]
\includegraphics{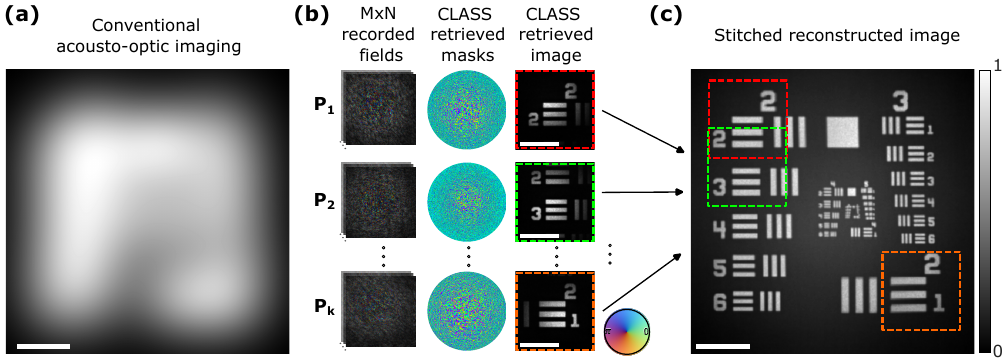}
\caption{\textbf{Numerical example and method concept} A numerically simulated object is scanned using $K=16$ probes smaller than the memory-effect range as illustrated Fig.~\ref{fig1}. Each tagged patch is illuminated with $M$=150 random illuminations. (a) Conventional AOI image is presented as a reference method where for each focused ultrasonic beam, we sum over all the field intensity, by that getting an acoustic resolution image (b) We retrieve $M$ fields on the detection diffuser plane for each patch. The CTR-CLASS method is applied to these fields to extract the effective distorting phase mask and the retrieved patch image. (c) In the final step of the reconstruction process, we stitch each patch image using cross-correlations into a wide FoV that extends beyond the memory-effect range. Scale bars, $50 \:\mu m$ }
\label{fig2}
\end{figure*}

\section{Methods}

\subsection{Concept}
Figure \ref{fig1} provides an illustration of the concept underlying our method, and the experimental setup used to realize it.

Our technique is based on acoustically modulating only a small segment of the target by focusing an ultrasound pulse to a spot (yellow circles in Fig.~\ref{fig1}a) that is smaller than the optical memory-effect range (green circle in Fig.1a). By holographically recording the acousto-optically tagged scattered light fields with a high-resolution camera via off-axis holography \cite{rosenfeld21,doktofsky20} (Fig.1b), we are able to estimate the effective reflection matrix from the camera plane to the acoustic focus and back, from a relatively small number of unknown random speckle illuminations (Fig.1b) via CTR-CLASS \cite{lee22}. 

Given that the acoustically-tagged region lies within the optical memory-effect range (Fig.1a), the scattering from the tagged target points to the camera is effectively isoplanatic, and can be represented by a single phase mask. As a result, applying the CLASS (or distortion matrix) algorithm on this matrix enables the retrieval of the effective scattering phase mask, and the diffraction-limited reconstruction of the object. 
This process can be repeated for numerous ($K$) parts of the object by methodically moving the ultrasound spot and measuring a matrix at each location. After obtaining images of all $K$ parts of the object, we stitch them in a straightforward manner into a single composite image. This allows us to achieve a large FoV image, providing a comprehensive view of the subject under study.

The approach can be realized using the conventional holographic acousto-optic setup such as the one illustrated in Fig.~\ref{fig1} (b). In this setup, a pulsed laser beam at an optical frequency $f_{opt}$ illuminates the sample, after passing through a controlled rotating diffuser, designed to generate random uncorrelated illuminations (Fig.1b, green). An ultrasound pulse (Fig.1b, yellow) at a central frequency $f_{US}$ is focused to a controlled location on the target.
The light passing through the acoustic focus is frequency shifted by the ultrasound pulse to a frequency ${f_{AO}}$ = ${f_{opt}}$ + ${f_{US}}$ (Fig.1b, light blue). The frequency-shifted complex-valued light field amplitude and phase are separated from the unmodulated light and captured by a high-resolution camera by off-axis phase-shifting interferometry \cite{rosenfeld21, doktofsky20,gross05}, utilizing a frequency-shifted reference arm. For each ultrasound focus position, $M$ random-illumination scattered light fields are measured, from which, we compute the effective reflection matrix and reconstruct the modulated part of the target, and the scattering phase-function via CTR-CLASS as explained below.

\subsection{CTR-CLASS}

Following the work of Lee et al. \cite{lee22}, we consider $m=1...M$ measured scattered light fields, $\vec{A}_m$ obtained under random uncorrelated illuminations, $\vec{S}_m$. By representing the optical transmission matrix from the illumination plane to the target plane as $\bm{P}_{ill}$, and the transmission matrix from the target plane to the camera plane by $\bm{P}_{det}$, each measured light field can be written as $\vec{A}_m = \bm{P}_{det}\bm{O} \bm{P}_{ill} \vec{S}_m$. The matrix $\bm{O}$ is the object transmission matrix. For a thin object, the matrix $\bm{O}$ is a diagonal matrix with the object transmission function along its diagonal.
By incorporating the $M$ measurements, $\vec{A}_m$, as columns in a matrix $\bm{A}$, the entire measured dataset can be represented in a matrix form as:
\begin{eqnarray}
\bm{A} = \bm{P}_{det}\bm{O} \bm{P}_{ill} \bm{S}
\label{eq:1}
\end{eqnarray}
where $\bm{S}$ is a matrix containing the input random patterns in its columns. By leveraging the fact that the illumination patterns are uncorrelated, i.e. $\bm{S}\bm{S}^\dagger = \bm{I + \epsilon}$ (where $\bm{I}$ is the identity matrix and $\bm{\epsilon}$ is the statistical noise matrix which diminishes as the number of realizations $M$ increases, eventually becoming negligible with a significant number of such uncorrelated illumination patterns), and the time-reversal symmetry of light propagation ($\bm{P}_{ill}\bm{P}_{ill}^{\dagger}=\bm{I}$), we can analyze the covariance matrix of $\bm{A}$: $\bm{R}=\bm{A}\bm{A}^{\dagger}$, as follows:
\begin{eqnarray}
\bm{R}=\bm{A}\bm{A}^\dagger = \bm{P}_{det}\bm{O}\bm{P}_{ill}\bm{S} \bm{S}^\dagger\bm{P}_{ill}^{\dagger} \bm{O}^\dagger \bm{P}^\dagger_{det} \nonumber \\ \approx \bm{P}_{det}\bm{O}_{eff} \bm{P}^\dagger_{det} = \bm{P}_{det}\bm{O}_{eff} (\bm{P}^{*}_{det})^T
\label{eq:2}
\end{eqnarray}

where $\bm{O}_{eff} \equiv |\bm{O}|^2$ (see Supplementary S1 for further explanations). 

Remarkably, as seen from Eq. 2, the covariance matrix $\bm{R}$ is mathematically analogous to the conventional reflection matrix from the camera plane to the object and back, when the target object is replaced by $\bm{O}_{eff}$. Hence, one can apply any conventional reflection-matrix analysis approach such as CLASS \cite{kang17,lee22} or distortion matrix \cite{badon2020distortion} on $\bm{R}$.

We chose to apply the recently introduced memory-efficient version of the CTR-CLASS algorithm, whose code is publicly available \cite{weinberg2023noninvasive}, since it allows the application of the CLASS algorithm directly on high pixel-count holographic datasets without the need to explicitly compute the memory-demanding reflection matrix $\bm{R}$

\section{Results}

\subsection{Numerical simulation}
To illustrate the efficacy of our approach in a clean and controlled scenario, we conducted a numerical simulation of the experimental configuration outlined in Fig.~\ref{fig1}. This setup was simulated by placing the object at a distance of $900 \: \mu m$ from two random phase masks positioned $30 \: \mu m$ apart from each other. The object was divided into $K=16$ patches, each sized below the isoplanatic patch size (Fig.~\ref{fig1} (a)). As a reference method, we presented the best achievable resolution using conventional AOI by scanning the object with a simulated acoustic probe continuously scanned over the target, where the image intensity in each position is obtained by summing the intensity of the modulated field (Fig.~\ref{fig2} (a)). Subsequently, we implemented our proposed approach. To this end, we subjected each ultrasonically-modulated area to 150 monochromatic ($\lambda = 700 \: nm$) random unknown illuminations, and then applied the CTR-CLASS algorithm to the simulated acquired fields. As a result, we were able to retrieve both the effective phase-mask distortion of each patch, and the corresponding patch image (Fig.~\ref{fig2}(b)). Following this process for all $K$ patches, the reconstructed $K$ images were mosaicked together in a straightforward manner, by simply summing them after relatively shifting each of them, resulting in a wide FoV image that significantly surpasses the memory-effect limitations (Fig.~\ref{fig2} (c)).

%%%%%%%%%%%%%%% Figure 3 - Experimental results%%%%%%%%%%%%%%%%%%%%%%
\begin{figure*}[t]
\includegraphics{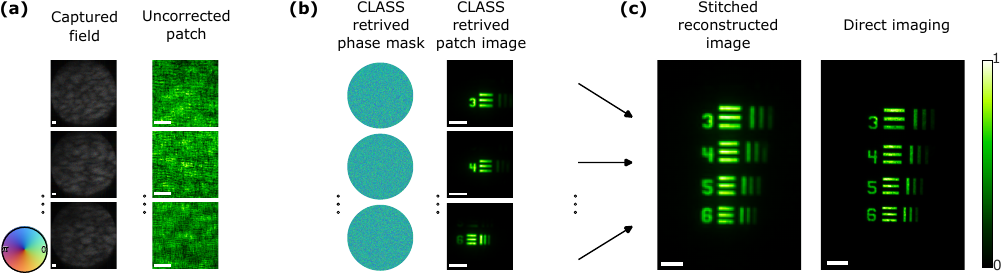}
\caption{\textbf{Experimental Imaging through optical diffusers beyond the optical memory-effect by acoustic-optic tagging}
Experimental image of $K=8$ tagged patches of a USAF resolution target hidden behind a scattering layer. (a) Each patch was illuminated with $M$=150 random illuminations by rotating the diffuser, each captured using holographic measurements, and as an initial, uncorrected measure, we utilized temporal STD on the field intensity in the conjugate object plane by taking the diagonal of the Fourier-transformed matrix of $\bm{R}$, which was significantly distorted due to scattering. (b) Subsequently, the CTR-CLASS algorithm was applied to correct the scattering distortion for each patch and retrieve the distortion phase mask. (c) Following correction, the individual patches were stitched into a composite object using cross-correlation. Additionally, we removed the detection patch diffusers to establish a ground truth comparison for the object and averaged the fields over 200 random illuminations.
Scale bars, $200 \:\mu m$}
\label{fig3}
\end{figure*}

\subsection{Experimental results}
Following the numerical affirmation of the approach, we performed an experimental demonstration using the experimental setup illustrated in Fig.~\ref{fig1} (b). The experimental setup is based on a nanosecond pulsed long-coherence laser with a wavelength of 532 $\mu m$ (Standa Ltd., Vilnius, Lithuania), a 25MHz ultrasound transducer (Olympus V324), and an sCMOS camera (Zyla 4.2 PLUS, Andor Technology, Belfast, UK). The experimental setup is identical to the experimental setup used in our previous work (Rosenfeld et al. \cite{rosenfeld21}), and its full technical details are provided and can be found in the Supplementary Section of Rosenfeld et al. \cite{rosenfeld21}.
In this proof of principle experimental demonstration, we focused on a specific segment of group 4 within a negative USAF-1951 resolution target submerged in water and placed between two scattering layers comprised of several $5 ^{\circ}$ scattering diffusers. We examined $K=8$ overlapping patches of the target, each probed by a focused ultrasound transducer operating at a central frequency of $f_{US}=25 MHz$, featuring ultrasound focus dimensions of $(D_x \times D_y = 149 \times 140 \: \mu m \times \mu m$) (FWHM).

For each of these $K$ regions the target was illuminated by the pulsed laser. Prior to reaching the sample, the laser beam passed through a controlled rotating diffuser, consisting of two 1° light-shaping diffusers from Newport Corp. (Irvine, CA, USA), which were rotated $M=150$ times for each patch, while the acoustic tagging process occurred for each of the $M$ illuminations. This process yielded a comprehensive collection of $K \times M$ fields, which were recorded using an sCMOS camera. The recording process involved utilizing a holographically frequency-shifted reference beam\cite{rosenfeld21}.
The recorded fields at the camera plane were digitally propagated to the scattering layer plane to apply the CTR-CLASS algorithm (See Supplementary S2). This digital conjugation allows to apply the computational phase correction at the correct plane, maximizing isoplanatism \cite{kwon2023computational}. 
We independently reconstructed the target image at each patch by applying the CTR-CLASS algorithm, while the uncorrected reconstruction, obtained by computational propagation to the object plane, and incoherent compounding, serves as a reference to highlight the correction effectiveness. 
The uncorrected incoherently-compounded reconstruction process is equivalent to the image reconstruction of dynamic speckle illumination \cite{ventalon2005quasi,ventalon2006dynamic}, and speckle SOFI \cite{dertinger2009fast, kim2015superresolution}. Interestingly, for distant objects, where the Fraunhofer approximation can be applied, this 'quasi-confocal' \cite{ventalon2005quasi} image can be also obtained from the diagonal of the Fourier-transformed matrix of $\bm{R}$ (Fig.~\ref{fig3} (a)) \cite{lee22,weinberg2023noninvasive}.

Following the $K$ reconstructions, we performed the final step of mosaicking the $K$ reconstructed images of the different patches into a full-field object image by simply summing the images after appropriately shifting each of them such that its cross-correlation with the neighboring patch image is maximized (Fig.~\ref{fig3} (b)). The final image is compared to an image obtained by direct imaging, achieved by removing the diffuser from the detection path and incoherently averaging 200 randomly illuminated realizations on the camera (Fig.~\ref{fig3} (c)).

%%%%%%%%%%%%%%% Figure 4 - Need of modulation %%%%%%%%%%%%%%%%%%%%%%
\begin{figure}[t]
\includegraphics{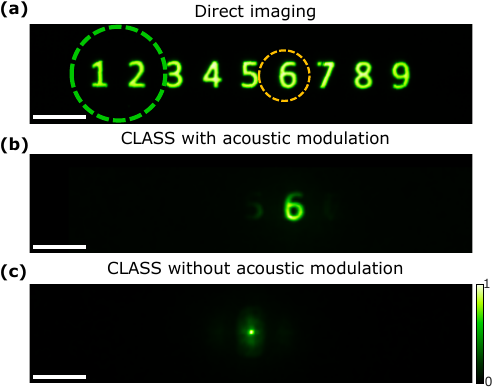}
\caption{\textbf{Effectiveness of acoustic modulation for extended objects beyond the memory-effect range} (a) An extended object located well beyond the memory-effect range (depicted in green) is imaged twice: once without acoustic modulation and the second time with acoustic probing (depicted in yellow) using a focused beam smaller than the isoplanatic patch size. In both cases, $M$=150 illuminations were employed. (b) When the CTR-CLASS algorithm is applied to the measurement without acoustic modulation, it converges to an incorrect result due to the 'anisoplanatic' distortion experienced by the object. (c) In contrast, when acoustic modulation is applied, filtering the field that did not pass through the probed area yields the correct result when applying the CTR-CLASS algorithm. Scale bars, $50 \:\mu m$}
\label{fig4}
\end{figure}

To complement this proof-of-principle, we conducted an additional experiment, where we demonstrated the critical role of acoustic modulation when imaging objects that extend significantly beyond the memory-effect range. 
In this experiment, we imaged a transmissive target featuring nine digits (Fig.~\ref{fig4} (a)), which spans an area $>3.5$ times larger than the memory-effect range of the scattering sample ($\Delta r_{mem} \approx 280 \:\mu m$). The critical importance of acoustic modulation was highlighted in two fashions: First, we demonstrate that when applying acoustic tagging, the algorithm effectively retrieves the tagged area (Fig.~\ref{fig4} (b)). This success is attributed to CLASS ability to effectively correct a single isoplanatic patch. Conversely, when the ultrasound field employing our method and simply capturing the field is not present, employing the CLASS algorithm on the captured field transmitted through the target proves ineffective, as expected (Fig.~\ref{fig4} (c)). This failure is due to the fact that the CLASS algorithm, as many state-of-the-art techniques, is based on the assumption of isoplanatic scattering, while the experimental scenario presents anisoplanatic distortions, as typical of deep tissue imaging.

\section{Discussion and Conclusions}

By employing localized acousto-optic modulation, we have successfully demonstrated that noninvasive imaging can be directly performed beyond the optical memory-effect range. Our approach necessitates only that the ultrasound focus size be within the scale of the memory-effect range, and the utilization of several tens to 100 random unknown illuminations.

While Rosenfeld et al. \cite{rosenfeld21} explored a similar advantage of acousto-optic tagging, our contribution lies in our utilization of well-established matrix-based algorithms, such as CLASS, allowing stable and interpretable reconstruction framework, as opposed to phase retrieval algorithms. This distinction offers two important advantages: first, we obtain an effective scattering phase mask that can be leveraged for optical diffraction-limited focusing, and second, we ensure computational tractability, making our method a practical and viable solution under practical settings.

A promising direction for future research involves integrating a Spatial Light Modulator(s) (SLMs) into the detection and/or illumination path, to allow physical focusing on the target, a step not taken in our current work. Incorporating an SLM is anticipated to exceed the focusing capabilities of previous methods that employed acousto-optic modulation, such as I-TRUE\cite{si2012breaking,ruan2014iterative}, TROVE\cite{judkewitz2013speckle}, and AOTM\cite{katz2019controlling}. Remarkably, the conventional AOTM\cite{katz2019controlling} or TROVE \ \cite{judkewitz2013speckle} experimental setups and data acquisition can be adapted to our proposed method, provided that the optical memory-effect is present.

Our proposed approach has three main limitations: The first is the necessity for the memory-effect range to align with the size of the ultrasound focus. The second limitation revolves around the need for distinct overlapping features between adjacent probed patches for successful image stitching using cross-correlations. 
Although this may seem restrictive, our control over the probed area alleviates this issue. 
A third challenge is the requirement for system and target temporal stability throughout the acquisition process, specifically in terms of the object's intensity transmission and the distortion of the medium between the object and the camera. A similar requirement exists in previous approaches such as TROVE and AOTM. 
However, it is important to note that the scattering in the illumination path, i.e., between the illumination source and the object, can be dynamic and does not need to remain static.

The number of measurements required for successful reconstruction is the same as in CTR-CLASS, and is studied in depth in Lee et al. \cite{lee22}. An empirical relationship between the number of measurements and the reconstruction quality, specifically for the experimental results data presented in Figure \ref{fig4}, is explored in Supplementary Materials S3.

An interesting avenue for future research involves exploring the possibility of jointly processing matrices of adjacent overlapping foci when the memory-effect is smaller than the acoustic focus size. Inspired by methods like AOTM and TROVE, this approach may offer the potential to overcome the current requirement for the size of the acoustic focus, by harnessing the joint information that exists in overlapping regions. This information is currently not utilized in the presented framework, where each acoustic foci data is separately analyzed. 

\section*{Supplementary Materials}
The supplementary materials provide further explanation and detailed mathematical derivations. Additionally, they include an empirical investigation that explores how the number of realizations affects image reconstruction.

\section*{Acknowledgements}
This work has received funding from the European Research Council under
the European Union’s Horizon 2020 Research and Innovation Program grant number 101002406.

\section*{Author Declarations}
\subsection*{Competing interests}
\noindent The authors declare no competing interests.
\subsection*{Author Contributions}
\noindent \textbf{Elad Sunray}: Formal analysis (lead); Software (lead); Visualization (lead); Writing – original draft (lead);  Writing – review \& editing (equal). Data curation (equal); Methodology (equal); Methodology (equal); . \textbf{Gil Weinberg}: Methodology (equal); Investigation (equal);  Data curation (equal); writing – review \& editing (equal). \textbf{Moriya Rosenfeld}: Investigation (equal); \textbf{Ori Katz}: Conceptualization (lead); Funding acquisition (lead); Supervision (lead); Project Administration (lead); Writing – review \& editing (equal).
\section*{Data availability}

The data that support the findings of this study are available from the corresponding author upon reasonable request.

\newpage
\onecolumngrid

\part*{\centering Supplementary Materials} 

\author{} %leave this blank
\setcounter{section}{0}
\setcounter{equation}{0}

% Beyond Memory-Effect Computational Adaptive Imaging via Acousto-optic scanning
\title{Beyond memory-effect matrix-based imaging in scattering media by acousto-optic gating - Supplementary Materials}

% Computational adaptive imaging beyond the optical memory effect via acousto-optic scanning gating
% Force line breaks with \\
\author{Elad Sunray}
 \affiliation{Institute of Applied Physics, Hebrew University of Jerusalem}%Lines break automatically or can be forced with \\

\author{Gil Weinberg}
\affiliation{Institute of Applied Physics, Hebrew University of Jerusalem}

\author{Moriya Rosenfeld}
 \affiliation{Institute of Applied Physics, Hebrew University of Jerusalem}

\author{Ori Katz}%
\affiliation{Institute of Applied Physics, Hebrew University of Jerusalem}
\email{orik@mail.huji.ac.il}

\date{\today}% It is always \today, today,

\maketitle
\section{CTR-CLASS}

The linear nature of a coherent imaging system and the wave equation makes choosing a basis consisting of independent illumination patterns possible. By writing the matrix of the system in this basis (the Green's function), we can fully describe the imaging process. Furthermore, if this matrix is pre-measured, one can obtain the output for any given illumination pattern by multiplying the matrix with the corresponding pattern \cite{popoff2010measuring}.
For instance, one possible choice for the basis is the set of point-illuminations, which aligns with the description of the imaging system in real space ($\vec{r}$-basis), such that using controlled illumination field $\vec{E}_{in} (\vec{r})$ the output field ${\vec{E}_{out}}(\vec{r})$ can be described as:
\begin{eqnarray}
\vec{E}_{out} = \bm{R}_{sys}\vec{E}_{in}
\label{eq:3}
\end{eqnarray}
Conceptually, we can decompose the system's matrix into a product of three distinct matrices, each corresponding to a distinct phase in the imaging process:

\begin{eqnarray}
\bm{R}_{sys} = \bm{P}_{det}\bm{O}\bm{P}_{ill}
\label{eq:4}
\end{eqnarray}
The first matrix $\bm{P}_{ill}$ represents the propagation through the optical system of the input field to the object plane. The second matrix $\bm{O}$ characterizes the interaction with the object, where this matrix is modeled as a diagonal matrix for thin objects, with the object reflection function $O(\vec{r})$ on its diagonal. The third matrix $\bm{P}_{det}$ governs the propagation of back from the object plane through the optical system to the detection. 

However, due to the very high number of modes in most common systems, measuring the reflection matrix can be daunting and time-consuming. In many practical imaging scenarios, such comprehensive measurements may be unnecessary for imaging within the optical memory effect. Recognizing this, the CTR framework \cite{lee22} presents an alternative approach that exploits the time-reversal symmetry of the illumination process for a phase-only distortion - which translates into the fact the matrix $\bm{P}_{ill}$, which represents the illumination step, is approximated as a unitary matrix $(\mathbf{{P}}_{ill}\mathbf{{P}}^\dagger_{ill} \approx \mathbf{I})$.
Instead of measuring the reflection matrix $\bm{R}_{sys}$ in an orderly set of modes, such as the point illumination basis or plane waves basis (in k-space), it is suggested to illuminate a set of random, uncorrelated illumination patterns.

The $M$ random illuminations used in the CTR methodology can be arranged in a matrix $\bm{S}$ in the $\vec{r}$-basis, where each illumination pattern corresponds to a matrix column. Due to their uncorrelated nature, the matrix \textbf{S} approximately satisfies the property $\bm{S} \bm{S}^\dagger = \mathbf{I} + \mathbf{\epsilon} \approx \mathbf{I}$, where $\bm{I}$ is the identity matrix and $\mathbf{\epsilon}$ represents the statistical noise matrix, which diminishes as the number of realizations $M$ increases and becomes negligible with a sufficient number of such realizations.

One can order the $M$ measurements in the columns of 'measurement matrix' $\bm{A}$ which can be written by:
\begin{eqnarray}
\bm{A} = \bm{R}_{sys}\bm{S} = \bm{P}_{det}\bm{O}\bm{P}_{ill}\bm{S}
\label{eq:5}
\end{eqnarray}
Then, rather than directly examining the reflection matrix $\bm{R}_{sys}$, the algorithm focuses on the matrix $\bm{R} \stackrel{\text{def}} = \bm{A}\bm{A}^\dagger$, which, from our analysis above can be written by:
\begin{eqnarray}
\bm{R}=\bm{A}\bm{A}^\dagger = \bm{P}_{det}\bm{O}\bm{P}_{ill}\bm{S} \bm{S}^\dagger\bm{P}_{ill}^\dagger \bm{O}^\dagger \bm{P}^\dagger_{det} \approx \bm{P}_{det}\bm{O}_{eff} \bm{P}^\dagger_{det} = \bm{P}_{det}\bm{O}_{eff} (\bm{P}^{*}_{det})^T
\label{eq:6}
\end{eqnarray}
where $\bm{O}_{eff} \stackrel{\text{def}} = \bm{O}\bm{O}^\dagger$ and has $|{O}(\vec{r})|^2$ on its diagonal, as $\bm{O}$ is diagonal matrix as we consider thin objects.

This matrix $\bm{R}$ has a similar structure to a full system reflection matrix in the real space ($\vec{r}$-basis), with the same illumination and detection transfer matrix up to phase conjugation.
In the context of imaging through a scattering medium within the memory-effect regime, we can further model the detection transfer matrix $ \bm{P}_{det}$ as a convolution Toeplitz matrix that represents the distorted point spread function (PSF) any point on the object is convoluted with.

The CTR-CLASS algorithm takes advantage of this unique structure of matrix $\bm{R}$ and estimates $\bm{P}_{det}$ and $\bm{O}_{eff}$ using the CLASS algorithm \cite{kang17}. This is achieved by iteratively leveraging the shifted inter-column correlations within $\bm{R}$.
A recent improvement introduced a memory-efficient version of the CTR-CLASS algorithm, significantly reducing memory usage to $O(MN)$, where $N$ denotes the number of camera pixels \cite{weinberg2023noninvasive}. This advancement enables the application of the algorithm to megapixel-scale images without the need to explicitly compute $\bm{R}$, which is sized $O(N^2)$.
It's important to note that while any matrix-based scattering compensation method could theoretically be applied given $\bm{R}$, in this work, we specifically utilize the CLASS algorithm.

\newpage
\section{CTR-CLASS correction in a conjugated plane}
As described above, the matrix $\bm{R}$ can be looked at as a reflection matrix for a system:
\begin{eqnarray}
\bm{R} = \bm{P}_{det}\bm{O}_{eff} (\bm{P}^{*}_{det})^T
\label{eq:7}
\end{eqnarray}

Where for an 'isoplanatic' patch where the distortion does not rely on object coordinates, the matrix $\bm{P}_{det}(\vec{r}_{diff},\vec{r}_{obj})$ describes a transfer matrix from each point on the detection scattering phase mask $\vec{r}_{diff}$ is transmitted towards the object plane with coordinates $\vec{r}_{obj}$. Hence, the effective field in the effective object plane is: $\vec{E}_{obj} = \bm{P}_{det}\vec{E}_{in}$. Assuming free-space configuration and that the distance $d$ between the object and the scattering layer is large enough to apply the Fresnel diffraction approximation, we can derive given an input field ${E}_{in}(x,y)$ the field at the object plane:
\begin{eqnarray}
{E}_{obj}(x,y)= \frac{e^{\frac{i2 \pi d}{\lambda}}}{i\lambda d} \iint {E}_{in}(x',y') 
e^{\frac{-i\pi}{\lambda d} [ (x-x')^2+(y-y')^2 ] } dx'dy'
\label{eq:8}
\end{eqnarray}
where $\lambda$ is the field wavelength. Thus, in a similar way to Choi et al. \cite{choi2022flexible}, each element of $\bm{R}(\vec{r}_{in},\vec{r}_{out})$ can be described as an input field of a point-illumination at $\vec{r}_{in}$ from the diffuser plane and point-detection $\vec{r}_{out}$ at the diffuser plane:
\begin{eqnarray}
{E}_{in}(x,y)= e^{-i\phi(x_{in},y_{in})}\delta(x-x_{in},y-y_{in})
\label{eq:9}
\end{eqnarray}
Where $\phi(x_{in},y_{in})$ is the phase distortion of the effective phase-mask and the sign is conjugated from Eq.~\ref{eq:7}. Using Eq.~\ref{eq:9} we obtain:
\begin{eqnarray}
{E}_{obj}(x,y) \propto  e^{-i\phi(x_{in},y_{in})}
e^{\frac{-i\pi}{\lambda d} [(x-x_{in})^2+(y-y_{in})^2 ]}
\label{eq:10}
\end{eqnarray}
Upon reflection from the effective object $O_{eff}$, the detection transfer matrix is applied:
\begin{align}
& {E}_{out}(x_{out},y_{out}) = e^{i\phi(x_{out},y_{out})} \frac{e^{\frac{i2\pi d}{\lambda}}}{i\lambda d} \iint {E}_{obj}(x',y')O_{eff}(x',y') 
e^{\frac{i\pi}{\lambda d} [ (x_{out}-x')^2+(y_{out}-y')^2 ] } dx'dy'  \nonumber &\\ 
&\propto e^{i\phi(x_{out},y_{out})} \iint e^{\frac{-i\pi}{\lambda d} [(x'-x_{in})^2+(y'-y_{in})^2 ]} O_{eff}(x',y') 
e^{\frac{i\pi}{\lambda d} [ (x_{out}-x')^2+(y_{out}-y')^2 ] } dx'dy'
 \nonumber &\\ 
&= e^{i[\phi(x_{out},y_{out})-\phi(x_{in},y_{in})]}e^{\frac{i \pi}{\lambda d}(x^2_{out}+y^2_{out}-x^2_{in}-y^2_{in})} \iint O_{eff}(x',y')e^{-i\frac{2 \pi}{\lambda d}[(x'(x_{out}-x_{in})+y'(y_{out}-y_{in})]}dx'dy'\nonumber &\\ 
&\equiv e^{i\phi_{eff}(x_{out},y_{out})}\tilde{O}_{eff}(\frac{x_{out}-x_{in}}{\lambda d},\frac{y_{out}-y_{in}}{\lambda d}) e^{-i\phi_{eff}(x_{in},y_{in})}&
\label{eq:11}
\end{align}

Where $\phi_{eff}(x,y) \stackrel{\text{def}} = \phi(x,y) + \frac{\pi}{\lambda d}(x^2+y^2)$ and $\tilde{O}_{eff}(\nu_x,\nu_y)$ is the Fourier spectrum of ${O}_{eff}$, leading to a scaled ${O}_{eff}$ with a scaling factor of ${\lambda d}$. 

So $\bm{R}(\vec{r}_{in},\vec{r}_{out})$ has a structure of a matrix similar to an isoplanatic imaging object beyond scattering layers system reflection matrix in the $\vec{k}$-basis. This matrix format aligns with the input requirements for the CLASS algorithm \cite{kang17,lee22}.

\newpage
\section{Reconstruction quality as a function of number measurements}

Despite the asymptotic requirement for the number of measurements being logarithmic relative to the number of degrees of freedom (i.e., the number of pixels in a holographic image) \cite{lee22}, the precise number of necessary measurements fluctuates based on the complexity of the object and distortion as well as the measurement signal-to-noise ratio (SNR). We empirically demonstrate in Figure S1 the relationship between the number of measurements, denoted as $M$, and the reconstruction quality. This figure presents the results of applying the CLASS algorithm to various numbers of measurements derived from the data illustrated in Figure 4 from the main text, showcasing how reconstruction quality improves with an increase in the number of measurements:

%%%%%%%%%%%%%%% Figure S1 - Different Ms %%%%%%%%%%%%%%%%%%%%%%
\begin{figure*}[htb!]
\includegraphics
{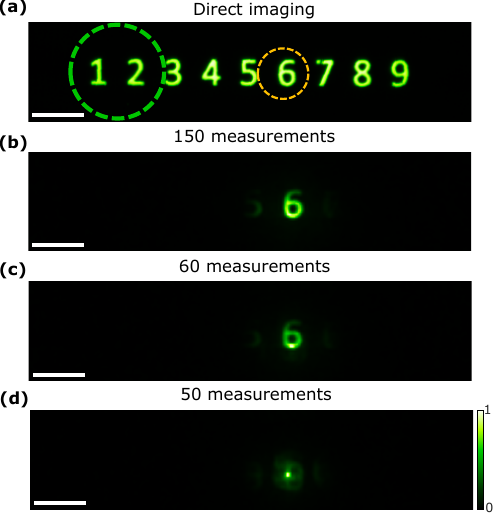}
        \renewcommand{\thefigure}{S1}
\caption{\textbf{Empirical impact of the number of measurements on the reconstruction quality}. (a) The same object presented in Fig.~4 from the main text, significantly bigger than the memory-effect range (in green), imaged with acoustic probing (in yellow), using a focused beam narrower than the 'isoplanatic' patch. (b) Displays the reconstruction outcome, identical to Fig. 4 from the main text, achieved with $M=150$ measurements for the specific patch. (c) The reconstruction is presented with $M=60$ measurements, where a small degradation in quality becomes evident. (d) Shows reconstruction with $M=50$ measurements, at which point the CTR-CLASS algorithm struggles to reconstruct. Scale bars, $50 \:\mu m$}
\label{figS1}
\end{figure*}

\newpage
\bibliography{Main}

\end{document}